\def\year{2021}\relax
\documentclass[letterpaper]{article} 
\usepackage{booktabs}
\usepackage{aaai21}  
\usepackage{times}  
\usepackage{helvet} 
\usepackage{courier}  
\usepackage[hyphens]{url}  
\usepackage{graphicx} 
\usepackage{natbib}  
\usepackage{caption} 
\title{Federated Crowdsensing: Framework and Challenges}
\author{
	Leye Wang\textsuperscript{\rm 1,2}, Han Yu\textsuperscript{\rm 3},  Xiao Han\textsuperscript{\rm 4}\\
}
\affiliations{
	\textsuperscript{\rm 1} Department of Computer Science and Technology, EECS, Peking University, China\\
	\textsuperscript{\rm 2} Key Laboratory of High Confidence Software Technologies (Peking University), Ministry of Education, China\\
	\textsuperscript{\rm 3} School of Computer Science and Engineering, Nanyang Technological University, Singapore\\
	\textsuperscript{\rm 4} School of Information Management and Engineering, Shanghai University of Finance and Economics, China
	
	leyewang@pku.edu.cn, han.yu@ntu.edu.sg, xiaohan@shufe.edu.cn
	
}
\begin{document}

\maketitle

\begin{abstract}
Crowdsensing is a promising sensing paradigm for smart city applications (e.g., traffic and environment monitoring) with the prevalence of smart mobile devices and advanced network infrastructure. Meanwhile, as tasks are performed by individuals, privacy protection is one of the key issues in crowdsensing systems. Traditionally, to alleviate users' privacy concerns, noises are added to participants' sensitive data (e.g., participants' locations) through techniques such as differential privacy. However, this inevitably results in quality loss to the crowdsensing task. Recently, federated learning paradigm has been proposed, which aims to achieve privacy preservation in machine learning while ensuring that the learning quality suffers little or no loss. Inspired by the federated learning paradigm, this article studies how federated learning may benefit crowdsensing applications. In particular, we first propose a federated crowdsensing framework, which analyzes the privacy concerns of each crowdsensing stage (i.e., task creation, task assignment, task execution, and data aggregation) and discuss how federated learning techniques may take effect. Finally, we summarize key challenges and opportunities in federated crowdsensing.

\end{abstract}

\section{Introduction}

Crowdsensing, by leveraging crowd participants to perform sensing tasks with today's powerful smart mobile devices (e.g., smartphones), has become one of the promising paradigm for urban sensing \cite{Ganti2011MobileCC,Zhang20144W1HIM}. While crowdsensing has witnessed a variety of successful applications in reality \cite{Guo2015MobileCS}, with the emerging concerns and regulations on data privacy (e.g., GDPR), crowdsensing community also puts more and more efforts into privacy protection for participants \cite{VergaraLaurens2017PrivacyPreservingMF}.

Take one of the representative sensitive data, participants' locations, as an example. Prior research has found that crowdsensing data can reveal participants' locations and trajectories \cite{Pournajaf2016ParticipantPI}, which may be used to infer their private information such as home and workplaces. State-of-the-art protection techniques usually achieve privacy protection through injecting imprecision (e.g., cloaking \cite{Pournajaf2014SpatialTA}) or inaccuracy (e.g., obfuscation like local differential privacy \cite{Wang2016DifferentialLP}) to perturb crowdsensing participants' locations. Nevertheless, \textbf{as these studies all need to modify participants' original locations, they would inevitably result in quality loss to crowdsensing}, which may be happened in the task assignment stage (e.g., participants need more time to finish a task \cite{Wang2017LocationPT,To2018PrivacyPreservingOT}), data aggregation stage (e.g., participants' sensed data cannot refer to a fine-grained location \cite{Wang2020SparseMC}), etc.

Recently, the \textit{federated learning} (FL) paradigm is proposed \cite{konevcny2016federated,yang2019federated}, which leverages distributed learning and encryption techniques to realize privacy-preserving model training with little loss (or even no loss) of model performance \cite{webank2018federated}:
\begin{equation}
	|V_\textit{fed} - V_\textit{cen}| < \delta
\end{equation}
where $V_\textit{fed}$ is the model performance obtained by federated learning where no user's private data are leaked, and $V_\textit{cen}$ is the model performance obtained by traditional learning where all the user's data are collected in a central server; $\delta$ is a bounded small positive number. Unlike prior privacy-preserving methods which mainly add noise into data for protection, FL targets at learning a lossless model by offloading computation processes involving sensitive private data to user clients. Then, only encrypted intermediate results (e.g., gradients) will be transmitted from clients to the server (or between clients) for constructing the final model.

Inspired by FL, in this paper, we consider a novel problem: \textbf{Is it possible to leverage FL in crowdsensing so as to minimize quality loss?} To address this problem, we propose a federated crowdsensing framework. In this framework, we identify several design considerations that may need to be considered in applying FL into crowdsensing, and then suggest some potential techniques toward these issues. Our federated crowdsensing framework follows the four-stage life cycle of crowdsensing identified by \citet{Zhang20144W1HIM}, thus including \textit{task creation}, \textit{task assignment}, \textit{task execution}, and \textit{data integration} stages. In particular, our framework enhances the four-stage life cycle by elaborating on the specific novel issues that need to be considered when federated learning is leveraged in crowdsensing. Besides, we also illustrate the potential techniques that may address these issues, and the future challenges and opportunities in federated crowdsensing.

\begin{figure*}
	\centering
	\includegraphics[width=.8\textwidth]{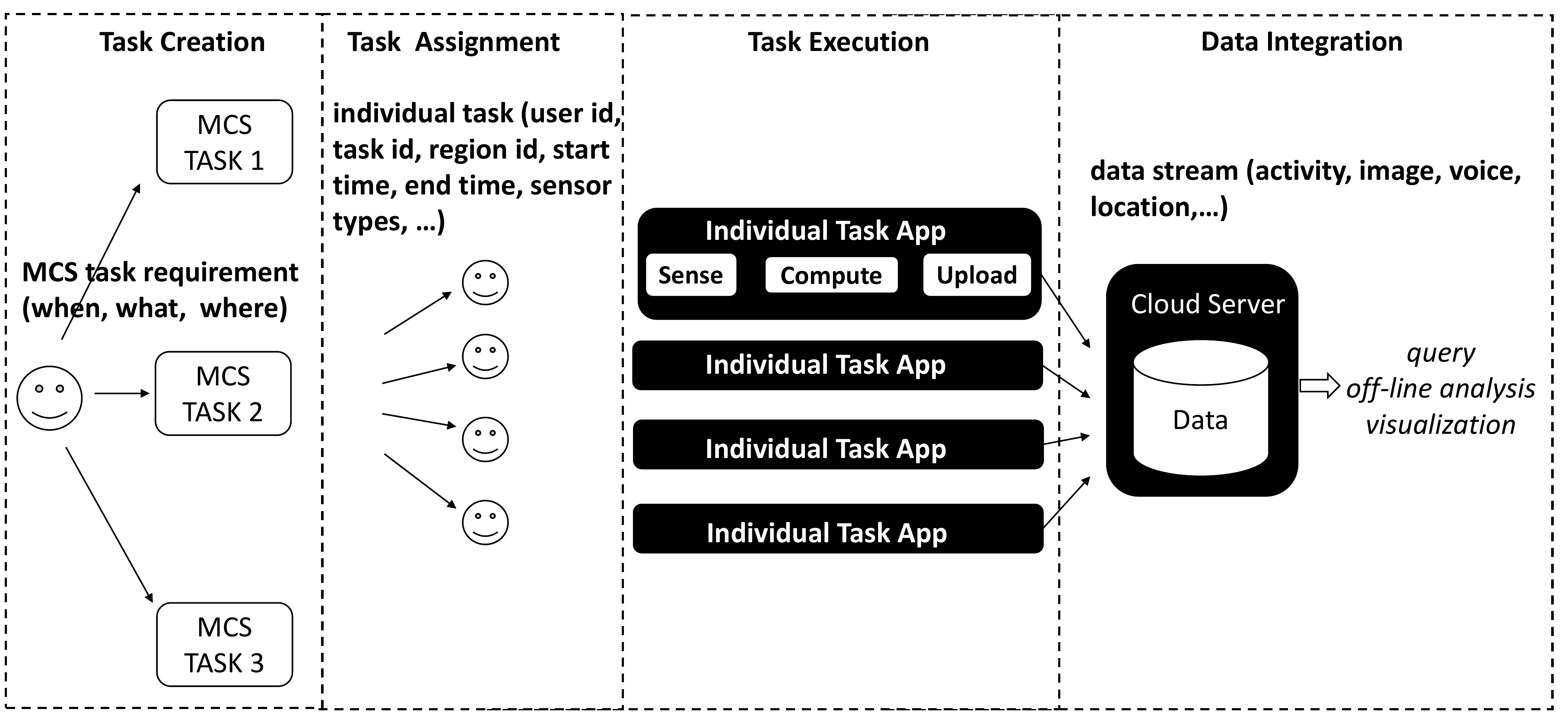}
	\vspace{-.5em}
	\caption{Four-stage life cycle crowdsensing framework \cite{Zhang20144W1HIM}.}
	\label{fig:four-stage}
	\vspace{-1em}
\end{figure*}

\section{Federated Crowdsensing Framework}

In this section, we introduce our federated crowdsensing framework by enhancing the four-stage life cycle framework \cite{Zhang20144W1HIM} with federated learning issues.

\subsection{Four-Stage Life Cycle of Crowdsensing}

Following \citet{Zhang20144W1HIM}, crowdsensing campaigns are usually conducted with a four-stage life cycle framework, including \textit{task creation}, \textit{task assignment}, \textit{task execution}, and \textit{data integration} (Figure~\ref{fig:four-stage}).

\begin{itemize}
	\item \textbf{Task Creation}: In the first stage, crowdsensing organizers or some task creators need to create the crowdsensing tasks. Generally, the task includes the time, location, task type (e.g., sensing noise, taking photos), data quality requirement (e.g., number of data samples), etc.
	
	\item \textbf{Task Assignment}: After tasks are created, the crowdsensing campaign needs to find enough participants for tasks. In general, there are two types of task assignment categories, server assigned task (SAT) and worker selected task (WST) \cite{Wang2017LocationPT}. WST lets every participant select their own preferred tasks, and the participants' private information is protected well. However, the pitfall is that the overall utility of the crowdsensing campaign would become lower, as each participant's local greedy selection may lead to bad overall utility. Today, SAT is more preferred in both academia and industry, as it can optimize the overall profits of the crowdsensing platform and the total welfare of participants better \cite{Wang2018TaskAI}. In this paper, we also focus on SAT.
	
	\item \textbf{Task Execution}: Task execution mainly refers that each participant conducts her sensing tasks locally. This stage is mostly related to clients' local data and computation, and the process is highly dependent on the specific crowdsensing task type. 
	
	\item \textbf{Data Integration}: After participants obtain their own sensed data, the final stage needs to integrate these data and generate the whole crowdsensing task results. This stage is also highly dependent on the crowdsensing task type. Meanwhile, different tasks may also share some common data integration steps, e.g., truth discovery \cite{Wang2014SurrogateMS} and missing data inference \cite{Wang2016SparseMC}.
\end{itemize}

\subsection{Federated Learning Issues for Crowdsensing}

With the four-stage life cycle, we elaborate on the federated learning issues that can be considered in each stage.

\subsubsection{Task Creation.}

In most cases, it seems no privacy leakage for task creation. For example, let us assume that a city government would like to run a noise crowdsensing campaign \cite{Rana2010EarphoneAE}. Then, the task information only includes the areas and time duration to sense. Such information will generally not incur privacy loss for any users.

Meanwhile, in some cases, tasks are created by some individual users and they may not be willing to reveal too much information about their tasks to the public, e.g., the detailed locations of the task \cite{To2018PrivacyPreservingOT}. For such tasks including private attributes, we thus need to design the crowdsensing system with the \textit{distributed attribute storage} functionality: some privacy-sensitive attributes of tasks will be stored in the task creators' clients instead of the crowdsensing server. To this end, distributed database technology should be taken into account \cite{zsu1990PrinciplesOD}.

\subsubsection{Task Assignment.}

In traditional crowdsensing task assignment, participants' useful information, such as historical trajectories, is often directly sent to the crowdsensing server and then the server aims to find the best matches between tasks and participants (e.g., minimize the travel distances between participants and tasks). However, in federated crowdsensing, users' private information like trajectories will not be sent to the server. Hence, in the federate learning setting, we need to do the task-participant matching by a distributed manner.

Regarding the scenario when all the task attributes are public, a natural way to do distributed matching is sending all the task information to participant clients and then the clients do the matching (e.g., calculating the distance between participants' current locations and task locations, and then returning the matching results). However, client-side distributed matching still suffers from certain concerns:

1) \textit{Privacy leakage from the matching results}. While participants can do task matching on their clients, it is still possible that the matching results returned back to the crowdsensing server may reveal some participants' private information. For example, if participants directly return the distance between their locations and task locations, we can easily infer a participant's location with more than three task distances by triangulation localization. Even if a participant only returns the matching results including nearby tasks (e.g., $< 1$ km) without detailed distance information, the server can still infer an approximate area where the participant is. To alleviate this issue, it is desirable to do the whole matching algorithm in a privacy-preserving manner and only the final task assignment results are revealed to the server. For example, to find a participant who is the nearest to a task, we can reply on \textit{multi-party secure ranking} algorithms. In this way, without revealing a participant's own private information (e.g., the distance to a specific task) to other participants or the server, we can rank all the participants according to the private information.

2) \textit{Malicious participants}. Offloading all the matching computation to the participants may lead to one potential risk: \textit{How to ensure that all the participants do their local task-participant matching trustfully?} For example, if certain participants want to be assigned with more tasks for higher incentives, they may mock their GPS locations. In this regard, the server should have some methods to check the validity of the participants' returned matching results. In this regard, we may need the techniques such as secure signature and public-key infrastructure \cite{Bonawitz2017PracticalSA}  to ensure that participants have trustfully follow our designed protocol to do task matching locally.

Furthermore, if task locations are stored in the task creators' side due to the privacy concerns, the task assignment stage will become even more challenging in federated crowdsensing. In other words, the task locations and participant locations are stored at the task creator and participant client sides, respectively; then, \textit{how to do task assignment by optimizing the overall utility (e.g., minimizing the total travel distance of the matched task-participant pairs) without leaking the locations of any tasks or participants}? To our knowledge, this is still an open research problem while we believe that encryption techniques such as homophonic encryption and multi-party secure computation would provide key tools to solve such problems. 

\begin{table*}[t]
	\footnotesize
	\centering
	\begin{tabular}{@{}lll@{}}
		\toprule
		& \textbf{Federated Crowdsensing Issues} & \textbf{Potential Techniques} \\ \midrule
		\textit{Task Creation} & private task attributes & distributed database \\\midrule
		\textit{Task Assignment} & \begin{tabular}[c]{@{}l@{}}1. distributed task-participant matching\\ 2. malicious participants\end{tabular} & \begin{tabular}[c]{@{}l@{}}secure multi-party computation\\ homophonic encryption\\ public key infrastructure\\ secure signature\\ ...\end{tabular} \\\midrule
		\textit{\begin{tabular}[c]{@{}l@{}}Task Execution \& \\ Data Integration\end{tabular}} & \begin{tabular}[c]{@{}l@{}}1. truth discovery\\ 2. missing data inference\\ 3. participant contribution assessment\\ 4. other task-dependent integration\end{tabular} & \begin{tabular}[c]{@{}l@{}}secure multi-party computation\\ homophonic encryption\\ federated matrix factorization\\ ...\end{tabular} \\ \bottomrule
	\end{tabular}
	\vspace{-.5em}
	\caption{Federated crowdsensing key issues and potential techniques.}
	\label{tbl:fedcrowd}
	\vspace{-1em}
\end{table*}

\subsubsection{Task Execution \& Data Integration.}

In traditional crowdsensing, participants' task execution usually only involves collecting sensed data. Then, the data are uploaded to the server for integration. However, in the federated crowdsensing setting, the raw sensed data would not be directly sent to the crowdsensing server for privacy protection. Hence, besides obtaining sensed data, the participant clients would need to perform more computations for data integration in a federated manner. In other words, not like traditional crowdsensing framework where task execution and data integration are split clearly (one in client side and one in server side), the task execution and data integration stages of federated crowdsensing are more mixed to each other as many computation processes now need to be conducted in the client sides for privacy protection. 

Take a common data integration procedure, truth discovery, as an example, which aims to find the sensing event truth from multiple (perhaps conflicted) sensed data from participants \cite{Wang2014SurrogateMS}. In traditional truth discovery mechanism, participants only need to send their sensed data to the server, and then the server would take over all the computation processes to discover the final truth sensed value. However, this may leak the participants' privacy, e.g., the server would know when and where a participant was located. To alleviate this privacy risk, running truth discovery in a federated manner without revealing participants' raw sensed data to the crowdsensing server becomes necessary. In fact, while not using the concept `federated learning', some pioneering research has also considered to conduct the privacy-preserving truth discovery in a distributed manner. For instance, \citet{Xu2019EfficientAP} leverage a secure aggregation protocol \cite{Bonawitz2017PracticalSA} based on the Shamir's secret sharing \cite{shamir1979share} to achieve privacy-preserving and robust truth discovery in crowdsensing when some participants may lose the network connections temporarily. Particularly, each participant uses secret sharing to transfer some pieces of their sensed data to other participants and conducts the truth discovery computation. In this way, participants' data would be leaked only if at least $t$ (a security threshold) participants collude with each other \cite{Xu2019EfficientAP}.

Besides truth discovery, there are other common crowdsensing data integration algorithms, such as missing data inference \cite{Wang2016SparseMC}, which need to be further investigated under the federated learning setting. For example, compressive sensing and matrix factorization has been widely used in missing data inference for urban monitoring crowdsensing tasks such as air pollution and temperature sensing \cite{Wang2015CCSTAQO,Rana2010EarphoneAE,Zhou2018LocationPD}. Recently, researchers have started exploring federated algorithms for matrix factorization \cite{chai2020secure}, which may serve as a good starting point to build crowdsensing federated data inference mechanisms.  

In addition, evaluating participants' contribution is an important issue after data integration, as it is often related to the crowdsensing management, such as incentive mechanism design and future participant selection \cite{Peng2018DataQG}. While participants do not upload sensed data directly to the server in federated crowdsensing, how to assess participants' data quality and contribution becomes more challenging. While how to compute participants' contribution is dependent on the task type and data integration mechanism, conducting participants' contribution assessment in a federated manner still needs much dedicated research.

\subsection{Summary}

In this section, we enrich the original four-stage crowdsensing framework to federated crowdsensing by elaborating on the issues to consider for federated crowdsensing. Table~\ref{tbl:fedcrowd} summarizes the key different issues between federated crowdsensing and the original four-stage framework. Moreover, we illustrate the possible techniques that can help address these issues for readers' reference.

Specifically, we believe that the most important stages that need to be carefully revised are \textit{Task Assignment} and \textit{Data Integration}.

\begin{itemize}
	\item \textbf{Task Assignment}: For task assignment, since participants' private information such as current locations cannot be directly sent to the server, some distributed task-participant matching algorithms need to be designed. Furthermore, this is also related to the \textit{Task Creation} stage, as task creators may not directly reveal the task locations due to privacy concerns, which makes federated task assignment even challenging.
	
	\item \textbf{Data Integration}: For data integration, in federated crowdsensing, we need to offload much more computation to participants' local clients so as to not reveal participants' sensed data. The data integration stage thus becomes more closely related to the \textit{Task Execution}  stage, since sensing and computation both need to be conducted in clients. 
\end{itemize}




\section{Challenges and Opportunities}

While the previous section has analyzed the federated learning issues that may emerge for each crowdsensing life cycle stage, in this section, we list more challenges that may cover different stages when we try to implement federated learning algorithms in crowdsensing.

\subsection{Energy Consumption in Federated Crowdsensing}

While nowadays mobile clients' computation capability is increasing rapidly, battery consumption is still a very serious concern for users' daily usage. For example, most flagship smartphones today, like \textit{iPhone 12}, can usually last for only around 10-hour usage without charging\footnote{https://www.apple.com/iphone-12-pro/specs/}. As federated crowdsensing will inevitably offload more computation burdens to the participants' client devices, the energy consumption incurred by federated learning thus needs to be investigated comprehensively. In literature, many smartphone programs with high computation overhead may be configured to run only when the smartphone is charging \cite{Xu2018DeepTypeOD}. However, most crowdsensing tasks have temporal deadlines, and whether waiting for charging is a valid option also needs to be carefully studied considering people's smartphone charging habits. 

\subsection{Network Connection in Federated Crowdsensing}

Another specialized issue of federated crowdsensing compared to other federated learning applications is the network connection instability. As crowdsensing participants usually move around a city, their network connections are often unstable. Then, how to ensure that the whole federated learning process can work even when some participant clients may lose network connections becomes a very important issue to ensure the robustness of federated crowdsensing systems. Some pioneering work on this has been conducted for some crowdsensing applications \cite{Xu2019EfficientAP}, while more work needs to be done to build a full life-cycle federated crowdsensing system. 

\subsection{Transfer Learning in Federated Crowdsensing}

In this article, we primarily discuss \textit{one} federated crowdsensing campaign. Actually, recent research has pointed out that multi-task and transfer learning may benefit the crowdsensing campaigns \cite{Wang2017SPACETACT,Wang2019CrossCityTL}. In federated crowdsensing, multi-task and transfer learning would become more challenging since data cannot be directly transferred between tasks. With the state-of-the-art development of federated multi-task and transfer learning algorithms \cite{Liu2020ASF,Smith2017FederatedML}, we believe that their applications into federated crowdsensing would also be a promising future research direction.

\section{Conclusion}

In this paper, we present a federated crowdsensing framework by incorporating federated learning techniques into crowdsensing for privacy protection and data regulation. Particularly, based on the four-stage life cycle crowdsensing paradigm, we enhance with the federated learning issues that need to be considered carefully. In addition, we present some specific research challenges and opportunities of federated crowdsensing. We expect that this article can attract more researchers into the federated crowdsensing area to achieve privacy-preserving, efficient, and robust crowdsensing campaigns. 

\bibliography{fedcrowd}

\end{document}